\documentstyle[titlepage,12pt]{article}

\begin{document}
\title{Stable and Unstable Circular Strings in Inflationary Universes}
\author{A.L. Larsen\thanks{E-mail: allarsen@nbivax.nbi.dk}\\
Nordita, Blegdamsvej 17, DK-2100 Copenhagen \O}
\maketitle
\begin{abstract}
It was shown by Garriga and Vilenkin that the circular shape of nucleated
cosmic strings, of zero loop-energy in de Sitter space, is stable in the sense
that the ratio of the mean fluctuation amplitude to the loop radius is
constant. This result can be generalized to all expanding strings (of
non-zero loop-energy) in de
Sitter space. In other curved spacetimes the situation, however, may be
different.

In this paper we develop a general formalism treating fluctuations around
circular strings embedded in
arbitrary spatially flat FRW spacetimes. As examples we consider Minkowski
space, de Sitter space and power law expanding universes.
In the special case of power law
inflation we find that in certain cases the fluctuations grow
much slower than the radius of the underlying unperturbed circular string.
The inflation of the universe thus tends to wash out the fluctuations and
to stabilize these strings.
\end{abstract}
\section{Introduction}
It is generally believed that cosmic strings can be formed in
the early history of the universe. This may take place
via the Kibble mechanism (see for instance Ref.\cite{kib}) before, in
or after the inflationary era and/or via
spontaneous quantum nucleation \cite{bas} during inflation. After
formation the evolution
(and fate) of the cosmic strings depend on the further evolution of the
universe and on the specific qualitative and quantitative features of the
strings themselves. Since cosmic strings may be responsible for galaxy
formation it is important to understand the evolution of strings through
the inflationary era and into the more "stable" radiation dominated
universe. This problem is conveniently discussed in the framework of
"test-strings" in fixed spatially flat Friedman-Robertson-Walker (FRW)
spacetimes, neglecting backreactions from the strings on the universe.

The main obstacle is the highly non-linear nature of the equations of motion
for a string in a curved background. As a consequence one often has to rely
on numerical calculations to obtain qualitative results. There are exceptions,
however. In some highly symmetric curved spacetimes it is possible to solve
the equations of motion analytically if a suitable ansatz for the string
(shape) is made. As examples we mention stationary strings in stationary
spacetimes and oscillating circular strings in axially symmetric spacetimes.

The purpose of the present paper is to consider small fluctuations around
circular strings in spatially flat FRW spacetimes. The evolution of the
fluctuations as compared to the evolution of the unperturbed underlying
circular string, we believe, is relevant for the evolution of closed
strings in general. In some cases the strings will turn out to be stable,
while in others they will eventually collapse into black holes or
disintegrate into smaller closed strings.

Our motivation is, besides the possible implications in a cosmological
scenario as discussed above, the recent interest in the stability
properties of cosmic and fundamental strings in curved spacetimes. Among
a variety of papers we mention the followings: Loust\'{o} and S\'{a}nchez
\cite{san} discussed the string-instabilities by considering first order
fluctuations around the center of mass of a small (fundamental) string in the
background of a black hole. A
similar analysis was carried out by the author \cite{all} for a large
(cosmic) string winding around a black hole in the equatorial plane.
String-instabilities in de Sitter space were discussed by Garriga and
Vilenkin \cite{gar1,gar2,gar3}, concentrating on the circular zero loop-energy
nucleated strings \cite{bas}. The analogue problem for circular strings
of non-zero loop-energy was considered in Ref.\cite{all}, concentrating
on the question of boundedness/unboundedness of the comoving fluctuations
in the two regions $r\rightarrow 0$ and $r\rightarrow\infty\;$ ($r$ is the
physical radius of the unperturbed circular string).
\vskip 6pt
\hspace*{-6mm}The
paper is organized as follows: in Section 2 we give the equation of
motion determining the physical radius of the circular string in the
FRW background, and we derive the linearized equations for the fluctuations
in the two directions perpendicular to the string world-sheet. In Section 3 we
first recall the results obtained in flat Minkowski space. We then discuss the
results for the
various types of expanding strings in de Sitter space and finally we consider
power law inflationary universes. Section 4 contains our conclusions.

Throughout the paper we use sign-conventions of Misner-Thorne-Wheeler
\cite{mis} and units where $G=1,\;c=1$ and the string tension
$(2\pi\alpha')^{-1}=1.$
\section{The Circular String Fluctuations}
In the zero thickness limit a cosmic string is described by the Nambu-Goto
action:
\begin{equation}
S=\int d\tau d\sigma\sqrt{-\det G_{AB}},
\end{equation}
where $G_{AB}$ is the induced metric on the world-sheet:
\begin{equation}
G_{AB}=g_{\mu\nu}x^\mu_{,A}x^\nu_{,B}.
\end{equation}
In this paper we are interested in configurations describing circular strings
with small fluctuations in a spatially flat FRW spacetime. The FRW line
element is usually written in the form:
\begin{equation}
ds^2=-dt^2+a^2(t)[dr^2+r^2 d\theta^2+r^2\sin^2\theta d\phi^2],
\end{equation}
where $t$ is the cosmic time and $a(t)$ is the scale factor.

For a circular string the physical radius is given by:
\begin{equation}
f(t)=r a(t)
\end{equation}
and it is therefore convenient, for our purposes, to write eq. (2.3) in terms
of $f$:
\begin{equation}
ds^2=-(1-H^2f^2)dt^2-2Hf\;dtdf+df^2+f^2d\theta^2+f^2\sin^2\theta d\phi^2,
\end{equation}
where we also introduced the Hubble function:
\begin{equation}
H(t)=\frac{1}{a(t)}\frac{da}{dt}.
\end{equation}
The unperturbed circular string with time dependent physical radius is obtained
by the ansatz:
\begin{equation}
t=\tau,\;\;\;f=f(\tau),\;\;\;\theta=\pi/2,\;\;\;\phi=\sigma.
\end{equation}
The physical radius $f(\tau)$ is to be determined by the equations of
motion corresponding to the action (2.1). Using eq. (2.2) with the spacetime
metric (2.5) and the ansatz (2.7) we find:
\begin{eqnarray}
\ddot{f}f\hspace*{-2mm}&-&\hspace*{-2mm}2Hf\dot{f}^3+
(6H^2f^2-1)\dot{f}^2+3Hf(1-2H^2f^2)\dot{f}\nonumber\\
\hspace*{-2mm}&-&\hspace*{-2mm}f^2\dot{H}+2H^4f^4-3H^2f^2+1=0,
\end{eqnarray}
where a dot denotes differentiation with respect to $t=\tau$. This equation
gives the physical string radius as a function of the cosmic time in a
spatially flat FRW spacetime \cite{akd,li,axe}. Since we are only interested in
timelike string evolution, we furthermore have the constraint:
\begin{equation}
G_{\tau\tau}=(\dot{f}-Hf)^2-1<0.
\end{equation}
The general solution of eq. (2.8), for arbitrary $H(\tau)$, does not seem to
be available. The complete solution is in fact only known in a few very
special cases like Minkowski space ($H=0$) and de Sitter space
($H=const.\equiv H_o\neq 0$), where the integrability is guaranteed by the
existence of the extra Killing vector $\partial/\partial_t$
(besides $\partial/\partial_\phi$)
in the metric (2.5). In many other cases it is however possible to find
special solutions and/or asymptotic solutions. We will return to that question
in Section 3, but for the moment we just assume that we have a solution
$f(\tau)$ for a spacetime $H(\tau)$ and we now address the question of
small fluctuations around it.

A covariant formalism of "linearized" fluctuations around arbitrary string
configurations in arbitrary curved spacetime backgrounds was developed by
Frolov and the author in Ref.\cite{fro} (see also Refs.\cite{car,guv}). From
the
two world-sheet tangent vectors:
\begin{equation}
\dot{x}^\mu=(1,\;\dot{f},\;0,\;0),\;\;\;\;x'^\mu=(0,\;0,\;0,\;1),
\end{equation}
we introduce two vectors $(n^\mu_\parallel,\;n^\mu_\perp)$, perpendicular to
the string world-sheet:
\begin{equation}
n^\mu_\perp=(0,\;0,\;\frac{1}{f},\;0),
\end{equation}
\begin{equation}
n^\mu_\parallel=\frac{1}{\sqrt{1-(\dot{f}-Hf)^2}}(\dot{f}-Hf,\;1+Hf
(\dot{f}-Hf),\;0,\;0),
\end{equation}
fulfilling ($R,S$ takes the values $"\perp"$ and $"\parallel"$):
\begin{equation}
g_{\mu\nu}n^\mu_R n^\nu_S=\delta_{RS},\;\;\;\;g_{\mu\nu}n^\mu_R x^\nu_{,A}=0,
\end{equation}
as well as the completeness relation:
\begin{equation}
g^{\mu\nu}=G^{AB}x^\mu_{,A}x^\nu_{,B}+\delta^{RS}n^\mu_R n^\nu_S.
\end{equation}
The general physical fluctuation is then expressed as:
\begin{equation}
\delta x^\mu=n^\mu_\perp\delta x^\perp+n^\mu_\parallel\delta x^\parallel,
\end{equation}
where $\delta x^\perp$ and $\delta x^\parallel$ are the comoving
fluctuations perpendicular to the string plane and in the string plane,
respectively. To first order in $\delta x^\mu$ it can now be shown that the
comoving fluctuations fulfill the following matrix equation \cite{fro}:
\begin{eqnarray}
&\Box\delta x_R+2\mu_{RS}\;^A(\delta x^S)_{,A}+(\nabla_A\mu_{RS}\;^A)
\delta x^S-\mu_{RT}\;^A\mu_S\hspace*{.3mm}^T\hspace*{.3mm}_A\delta
x^S&\nonumber\\
&+\frac{2}{G^C\hspace*{.3mm}_C}\Omega_R\hspace*{.3mm}^{AB}
\Omega_{SAB}\delta x^S-
h^{AB}x^\mu_{,A}x^\nu_{,B}R_{\mu\rho\sigma\nu}n^\rho_R n^\sigma_S\delta x^S=0.&
\end{eqnarray}
Here $h_{AB}$ and $G_{AB}$ are the intrinsic (Polyakov) and induced metric,
respectively, which in our case are identical since we are working with the
Nambu-Goto action. $\Omega_{RAB}$ and $\mu_{RSA}$ are the second fundamental
form and normal fundamental form, respectively \cite{eis}:
\begin{equation}
\Omega_{RAB}=g_{\mu\nu}n^\mu_R x^\rho_{,A}\nabla_\rho x^\nu_{,B},
\end{equation}
\begin{equation}
\mu_{RSA}=g_{\mu\nu}n^\mu_R x^\rho_{,A}\nabla_\rho n^\nu_S,
\end{equation}
where $\nabla_\rho$ is the spacetime covariant derivative. $\Box$ and
$\nabla_A$ are the world-sheet d'Alambertian and covariant derivative,
respectively, determined by the world-sheet metric (2.2). Finally
$R_{\mu\rho\sigma\nu}$ is the Riemann tensor of the metric (2.5), and
since we are using a somewhat unconventional parametrization for the FRW
spacetime we have listed the non-vanishing components in the Appendix.

Generally eqs. (2.16) constitute an extremely complicated set of coupled
partial differential equations. Fortunately some simplifications arise for
the unperturbed circular strings under consideration here. Using the explicit
expressions for the tangent vectors and normal vectors (2.10)-(2.12), it is
easily seen that all components of the normal fundamental form vanish, while:
\begin{equation}
\Omega_{\perp AB}=0,\;\;\;\;\Omega_{\parallel AB}\Omega_\parallel\;^{AB}=
\frac{2}{f^2}\frac{(1+Hf(\dot{f}-Hf))^2}{1-(\dot{f}-Hf)^2}.
\end{equation}
The relevant projections of the Riemann tensor, appearing in eqs. (2.16) are
evaluated using the explicit expressions of $R_{\mu\rho\sigma\nu}$ given in
the Appendix. After a little algebra one finds:
\begin{equation}
G^{AB}x^\mu_{,A}x^\nu_{,B}R_{\mu\rho\sigma\nu}n^\rho_\perp n^\sigma_\perp=
\frac{\dot{H}}{(\dot{f}-Hf)^2-1}-2H^2,
\end{equation}
\begin{equation}
G^{AB}x^\mu_{,A}x^\nu_{,B}R_{\mu\rho\sigma\nu}n^\rho_\parallel
n^\sigma_\parallel=\frac{2(\dot{f}-Hf)^2-1}{1-(\dot{f}-Hf)^2}\dot{H}-2H^2,
\end{equation}
while the cross terms $(\propto\;n^\rho_\perp n^\sigma_\parallel)$ vanish.
We eventually end up with two separated partial differential equations in the
form:
\begin{equation}
\Box\delta x_\perp+V_\perp(\tau)\delta x_\perp=0,
\end{equation}
\begin{equation}
\Box\delta x_\parallel+V_\parallel(\tau)\delta x_\parallel=0,
\end{equation}
where:
\begin{equation}
V_\perp=\frac{\dot{H}}{1-(\dot{f}-Hf)^2}+2H^2,
\end{equation}
\begin{equation}
V_\parallel=\frac{1-2(\dot{f}-Hf)^2}{1-(\dot{f}-Hf)^2}\dot{H}+
\frac{2}{f^2}\frac{(1+Hf(\dot{f}-Hf))^2}{1-(\dot{f}-Hf)^2}+2H^2,
\end{equation}
and the d'Alambertian is given by:
\begin{eqnarray}
\Box\hspace*{-2mm}&=&\hspace*{-2mm}\frac{1}{\sqrt{-G}}\partial_A(\sqrt{-G}\;
G^{AB}\partial_B)\nonumber\\
\hspace*{-2mm}&=&\hspace*{-2mm}\frac{-\partial^2_\tau}
{1-(\dot{f}-Hf)^2}+\frac{\partial^2_\sigma}{f^2}-
H\frac{1-2(\dot{f}-Hf)^2}{1-
(\dot{f}-Hf)^2}\partial_\tau,
\end{eqnarray}
using eqs. (2.2), (2.7) as well as eq. (2.8) to eliminate the $\ddot{f}$-terms.

Since the potentials in eqs. (2.22)-(2.23) depend on $\tau$ only, it is
convenient to Fourier expand the comoving fluctuations:
\begin{equation}
\delta x_R(\tau,\sigma)=\tilde{\sum_{n\in Z}}C_{nR}(\tau)e^{-in\sigma},
\end{equation}
where $C_{nR}=C^\ast_{-nR}$ and the tilde denotes summation for $\mid n\mid
\neq 0,1$ only. The zero modes and the $\mid n\mid=1$ modes are excluded from
the summation since they do not correspond to "real" physical fluctuations on
a circular string [5-7]. They describe spacetime translations and
rotations, that do not change the circular shape of the string.

The Fourier expansions reduce eqs. (2.22)-(2.23) to ordinary second order
differential
equations determining $C_{nR}(\tau)$. In
the next section we consider these
equations in various cases of cosmological interest.
\section{Special Cases}
\setcounter{equation}{0}
In this section we use the general formalism of Section 2 in some special
cases.
As our first example we consider Minkowski space. This case has already been
discussed in the literature (Appendix A of Ref.\cite{gar2}), but we
include it here to get an independent
check of equations (2.22)-(2.23). We then consider expanding strings in
de Sitter space and as an example of
other families of inflationary backgrounds, we consider power law expanding
universes.
\subsection{Minkowski Space}
For Minkowski space the Hubble function equals zero and equation (2.8),
determining the radius of the unperturbed circular string reduces
to \cite{vil}:
\begin{equation}
\ddot{f}f-\dot{f}^2+1=0.
\end{equation}
The solution is just a trigonometric function:
\begin{equation}
f(\tau)=r_m\cos(\frac{\tau-\tau_o}{r_m}),
\end{equation}
where $r_m$ is the maximal radius obtained at $\tau=\tau_o.$ The string then
contracts and collapses at $\tau=\tau_o+\pi r_m/2$. We can now easily
calculate the fluctuations around the configuration (3.2). The Fourier
transformed versions of eqs. (2.22)-(2.23), using also eqs. (2.24)-(2.27),
reduce to:
\begin{equation}
\ddot{C}_{n\perp}+n^2 C_{n\perp}=0,
\end{equation}
\begin{equation}
\ddot{C}_{n\parallel}+(n^2-\frac{2}{\cos^2(\tau-\tau_o)})C_{n\parallel}=0,
\end{equation}
where we have rescaled the time parameter $\tau\rightarrow \tau r_m.$ These two
equations are solved by:
\begin{equation}
C_{n\perp}(\tau)=A_{n\perp}\cos n\tau+B_{n\perp}\sin n\tau,
\end{equation}
\begin{eqnarray}
C_{n\parallel}(\tau)\hspace*{-2mm}&=&\hspace*{-2mm}A_{n\parallel}[n\cos n(\tau-
\tau_o)+\tan(\tau-\tau_o)\sin n(\tau-\tau_o)]\nonumber\\
\hspace*{-2mm}&+&\hspace*{-2mm}B_{n\parallel}[n\sin n(\tau-\tau_o)-
\tan(\tau-\tau_o)\cos n(\tau-\tau_o)].
\end{eqnarray}
Not surprisingly the fluctuations in the direction perpendicular to the
plane of the string are completely finite and regular. For the fluctuations
in the plane of the string we see that they blow up for $\tau\rightarrow
\tau_o+\pi/2,$ that is when the unperturbed string collapses. Note that the
fluctuations (3.5)-(3.6) are the comoving fluctuations, i.e. the fluctuations
as seen by an observer travelling with the unperturbed circular string. For an
external observer the fluctuations (3.6) should be multiplied by the Lorentz
contraction factor $\sqrt{1-\dot{f}^2}\;$ (generally $\sqrt
{1-(\dot{f}-Hf)^2}\;$). In the present case this factor exactly cancels the
divergent terms in eq. (3.6), so that all
fluctuations in eqs. (3.5)-(3.6) are finite
and regular for an external observer. This is however still not the final
picture: to discuss the stability of the circular shape the relevant quantity
is the ratio of the fluctuation amplitude to the radius of the unperturbed
circular string. Since the radius of the string goes to zero during the
collapse, it follows that for suitable initial conditions this ratio
will blow up, and the string will develop cusps and probably disintegrate
into smaller loops. For other initial conditions the ratio will stay
approximately constant until the string falls into its own
Schwarzschild radius, and the string will collapse into a black hole in a
"regular" way. These possibilities were discussed in more detail in
Ref.\cite{gar2}.
\subsection{de Sitter Space}
For de Sitter space the Hubble function is constant $H=const.\equiv H_o\neq 0,$
so that eq. (2.8) has no explicit time dependence and is therefore
integrable. The complete solution is
actually known, and may be found in Ref.\cite{san2} (in Ref.\cite{san2} the
solution is written in terms of a time parameter related to the cosmic time by
elliptic theta-functions). The fluctuations around circular strings in de
Sitter space were discussed by the author in Ref.\cite{all}, using the
static parametrization and the conformal gauge. Here we shall give some more
results.
\vskip 6pt
\hspace*{-6mm}A circular string configuration in de Sitter space of special
interest is provided by the nucleated string of Basu, Guth and Vilenkin
\cite{bas}. The nucleated string has zero loop-energy and is explicitly given
by:
\begin{equation}
f(\tau)=\frac{1}{H_o}\sqrt{e^{2H_o(\tau-\tau_o)}+1},
\end{equation}
where $\tau_o$ is a constant, i.e. it expands
with the same rate as the universe (for $\tau-\tau_o>>1/H_o$).
The explicit form (3.7) was obtained using the instanton method \cite{bas}, but
it is easily shown that eq. (2.8) is fulfilled, as it should be. It is
convenient to write the solution in terms of the time parameter $\tau_c$:
\begin{equation}
\tau-\tau_o=\frac{1}{H_o}\log\tan(\tau_c-\tau_o),
\end{equation}
so that $\tau\in\;]-\infty,\;\infty[\;$ corresponds to $\tau_c\in\;]\tau_o,\;
\tau_o+\pi/2[\;.$ The solution (3.7) is now:
\begin{equation}
f(\tau_c)=\frac{1}{H_o}\frac{1}{\cos(\tau_c-\tau_o)}.
\end{equation}
The string expands from horizon size at $\tau_c=\tau_o$ towards infinity for
$\tau_c\rightarrow\tau_o+\pi/2.$ In the $(\tau_c,\sigma)$ coordinates the
Fourier transformed versions of eqs. (2.22)-(2.23) reduce to:
\begin{equation}
\frac{d^2 C_{nR}}{d\tau^2_c}+(n^2-\frac{2}{\cos^2(\tau_c-\tau_o)})C_{nR}=0.
\end{equation}
Two remarks are worth doing here: the two fluctuation equations take exactly
the same form. This is somewhat surprising since, although the de Sitter
spacetime is isotropic, the isotropy is broken by the plane circular string.
Secondly, the two equations take exactly the same form as the
$C_{n\parallel}$-equation in Minkowski space (3.4), so they
are both solved by eq. (3.6). The
physical interpretation is however completely different here. Eq. (3.4)
describes the fluctuations around a collapsing string in Minkowski space.
After Lorentz contraction we found that the ratio of the fluctuation amplitude
to the string radius blowed up, leading to unstable strings. Eq. (3.10), on
the other hand, describes the
fluctuations around an expanding string in de Sitter space. The Lorentz
contraction plays no role here (for $\tau-\tau_o>>1/H_o$), and by comparing
eqs. (3.6) and (3.9) we see that the ratio of the mean
fluctuation amplitude to the
string radius is constant (c.f. Ref.\cite{gar3}), indicating
that the string is stable.

The solution (3.7) is a very special one (zero loop-energy), so it is not
obvious whether the results for the fluctuations can be generalized to
other expanding circular strings in de Sitter space. The evolution of
arbitrary circular strings in de Sitter space was discussed in Ref.\cite{san2}.
An expanding circular string, that has passed the horizon from the inside,
expands with the same rate as the universe ($H_of>>1$):
\begin{equation}
f(\tau)\propto e^{H_o\tau}
\end{equation}
In this case the fluctuation equations become approximately:
\begin{equation}
\ddot{C}_{nR}+H_o\dot{C}_{nR}-2H_o^2C_{nR}\approx 0,
\end{equation}
with solutions:
\begin{equation}
C_{nR}(\tau)\approx A_{nR}e^{H_o\tau}+B_{nR}e^{-2H_o\tau}.
\end{equation}
So, due to the $A_{nR}$-terms, the fluctuations grow with the same rate as the
universe, and the ratio of the fluctuation amplitude to the radius of the
circular string (3.11) is constant for these strings also.

In de Sitter space the physical radius of the horizon is constant ($=1/H_o$)
so in principle there could also be expanding strings that never cross the
horizon. Such strings would expand much slower than the universe and would
therefore be completely different from the strings discussed above. This type
of circular strings has actually been found \cite{mik}:
\begin{equation}
f(\tilde{\tau})=\frac{1}{H_o\sqrt{2}}\tanh\frac{\tilde{\tau}}{\sqrt{2}},
\end{equation}
where the time parameter $\tilde{\tau}$ is introduced by:
\begin{equation}
t=\tau=\frac{1}{H_o}[\tilde{\tau}+\log(1-\frac{1}{\sqrt{2}}\tanh\frac
{\tilde{\tau}}{\sqrt{2}})].
\end{equation}
Notice that $f(0)=0\;$, $f(\infty)=1/(H_o\sqrt{2})$ and $df/d\tilde{\tau}>0$.
For this solution the fluctuation equations become:
\begin{equation}
\frac{d^2C_{n\perp}}{d\tilde{\tau}^2}+(n^2-\tanh^2\frac{\tilde{\tau}}
{\sqrt{2}})C_{n\perp}=0,
\end{equation}
\begin{equation}
\frac{d^2C_{n\parallel}}{d\tilde{\tau}^2}+(n^2-\tanh^2\frac{\tilde{\tau}}
{\sqrt{2}}-\coth^2\frac{\tilde{\tau}}{\sqrt{2}})C_{n\parallel}=0.
\end{equation}
These two equations can be solved explicitly in terms of hypergeometric
functions in the variable $z\equiv-\sinh^2\frac{\tilde{\tau}}{\sqrt{2}}.$
For our purposes it is however sufficient to consider the somewhat simpler
approximate solutions. Starting from $f=0$ at $\tilde{\tau}=0$ the string
size grows steeply towards the asymptotic size $f=1/(H_o\sqrt{2}).$ Then
eqs. (3.16)-(3.17) reduce to:
\begin{equation}
\frac{d^2C_{n\perp}}{d\tilde{\tau}^2}+(n^2-1)C_{n\perp}\approx 0,
\end{equation}
\begin{equation}
\frac{d^2C_{n\parallel}}{d\tilde{\tau}^2}+(n^2-2)C_{n\parallel}\approx 0.
\end{equation}
For the "shape-changing" modes $\mid n\mid\geq 2$ (c.f. the discussion
after eq. (2.27)) the solutions are simple trigonometric functions. It
follows that also in this case the ratio of the mean fluctuation amplitude to
the radius of the circular string is constant.
\subsection{Power Law Inflation}
In this subsection we consider power law inflationary universes. The scale
factor is given by:
\begin{equation}
a(\tau)=a_p\tau^p,
\end{equation}
where $a_p$ is a dimensionfull constant, which is introduced to ensure that
$a(\tau)$ is dimensionless. The dynamics of the unperturbed circular strings
in the backgrounds (3.20) was investigated in Ref.\cite{axe}, using both
numerical and analytical methods. For $0<p<1,$ which includes as special
cases matter and radiation dominated universes, the circular string
dynamics is very similar to Minkowski space: expanding strings reach a
maximal size after which they collapse. For $p>1,$ corresponding to the
inflationary universes, the dynamics is much more interesting, so we
restrict ourselves to that case in the following.

The Hubble function of the spacetime (3.20) is $H(\tau)=p/\tau$ and for $p>1$
there is an event horizon with physical radius:
\begin{equation}
f_{EH}(\tau)=\frac{\tau}{p-1}.
\end{equation}
Circular strings outside this horizon tend to expand with the same rate as the
universe ($f>>\tau$):
\begin{equation}
f(\tau)\propto \tau^p;\;\;\;\;p>1.
\end{equation}
This is similar to de Sitter space. For such strings the fluctuation
equations become:
\begin{equation}
\ddot{C}_{nR}+\frac{p}{\tau}\dot{C}_{nR}-\frac{p(2p-1)}{\tau^2}C_{nR}\approx 0,
\end{equation}
with solutions:
\begin{equation}
C_{nR}(\tau)\approx A_{nR}\tau^p+B_{nR}\tau^{1-2p}
\end{equation}
and the fluctuations grow with the same rate as the unperturbed strings, which
again grow with the same rate as the universe. These strings are therefore
stable in the same sense as the expanding strings in de Sitter space, c.f.
Subsection 3.2.

We can however find other types of expanding strings in the
spacetimes (3.20). In the power law inflationary universes the event horizon
size increases proportionally to $\tau,$ eq. (3.21). In
principle this allows the
existence of monotonically expanding circular strings, that never pass the
horizon from the inside. A string solution of this type is in fact provided
by:
\begin{equation}
f(\tau)=\frac{\tau}{\sqrt{2p(p-1)}};\;\;\;\;p>1.
\end{equation}
It is easily checked that this solution fulfills eqs. (2.8) and (2.9) as well
as:
\begin{equation}
\dot{f}>0,\;\;\;\;\;\;f(\tau)<f_{EH}(\tau).
\end{equation}
We note in passing that the solution (3.25) is actually well-defined for
$p<0$ also. In that case it is convenient to consider instead:
\begin{equation}
\hat{f}(\tau)=\frac{\tau_o-\tau}{\sqrt{2p(p-1)}};\;\;\;\;p<0
\end{equation}
which solves eq. (2.8) in the spacetime with scale factor:
\begin{equation}
\hat{a}(\tau)=\frac{a_p}{(\tau_o-\tau)^{\mid p\mid}};\;\;\;\;p<0
\end{equation}
In this case eq. (3.27) describes a collapsing string (for $\tau\rightarrow
\tau_o$) in a super-inflationary universe ($\dot{H}>0$) which blows up
for $\tau\rightarrow\tau_o.$

But let us return to the expanding string (3.25) in the power law inflationary
universe. The potentials (2.24)-(2.25) become:
\begin{equation}
V_\perp(\tau)=\frac{2p^3}{(p+1)\tau^2},\;\;\;\;\;\;V_\parallel(\tau)=
\frac{2p(2p^2-1)}{(p+1)\tau^2}
\end{equation}
and the Fourier transformed fluctuation equations take the form:
\begin{equation}
\ddot{C}_{n\perp}+\frac{1}{\tau}\dot{C}_{n\perp}+\frac{1}{\tau^2}
[n^2(p^2-1)-p^2]C_{n\perp}=0,
\end{equation}
\begin{equation}
\ddot{C}_{n\parallel}+\frac{1}{\tau}\dot{C}_{n\parallel}+\frac{1}{\tau^2}
[n^2(p^2-1)-2p^2+1]C_{n\parallel}=0.
\end{equation}
These two equations are solved by:
\begin{equation}
C_{n\perp}(\tau)=A_{n\perp}\tau^{\alpha_+}+B_{n\perp}\tau^{\alpha_-},
\end{equation}
\begin{equation}
C_{n\parallel}(\tau)=A_{n\parallel}\tau^{\beta_+}+B_{n\parallel}\tau^{\beta_-},
\end{equation}
where:
\begin{equation}
\alpha_\pm=\pm\sqrt{n^2(1-p^2)+p^2},
\end{equation}
\begin{equation}
\beta_\pm=\pm\sqrt{n^2(1-p^2)+2p^2-1}.
\end{equation}
Notice that for $p>1$ and $n\geq 2$ ( which is the case we are interested in,
c.f. the discussion after eq. (2.27)):
\begin{equation}
\alpha^2_\pm-1=(1-n^2)(p^2-1)<0,\;\;\;\;\beta^2_\pm-1=(2-n^2)(p^2-1)<0,
\end{equation}
so that:
\begin{equation}
\alpha^2_\pm<1,\;\;\;\;\;\;\beta^2_\pm<1.
\end{equation}
If $\alpha^2_\pm<0\;$ ($\alpha_\pm$ purely imaginary) the fluctuations (3.32)
are represented by regularly oscillating trigonometric functions in the form
$\;\sim \cos(\mid\hspace*{-2mm}\alpha_\pm\hspace*{-2mm}\mid
\hspace*{-1mm}\log\tau)\;.$ If
$0<\alpha^2_\pm<1\;$ ($\alpha_\pm$ real) the fluctuations grow with a power of
$\tau,$ but always strictly slower than $\tau.$ Finally if $\alpha^2_\pm=0\;$
(this is only possible for very special values of $p$) eq. (3.32) is not the
general solution of eq. (3.30). Instead we find that the fluctuations grow as
the logaritm of $\tau.$ In any case we find that the ratio of the
$C_{n\perp}$-fluctuation
amplitude to the radius of the unperturbed circular string (which grows
proportionally to $\tau,$ eq. (3.25)) is a decreasing function of $\tau.$ The
same conclusion is drawn for the fluctuations $C_{n\parallel}.$ This
behaviour is very different from the behaviour found in de Sitter space. For
the expanding string solutions (3.25) the $\mid n\mid\geq 2$ fluctuation modes
are washed out by the inflationary expansion of the universe. The circular
shape thus becomes more and more stable in this case.
\section{Conclusion}
In conclusion we have studied small fluctuations around circular strings in
spatially flat FRW spacetimes. Our main results are the two separated second
order differential equations (2.22)-(2.23) determining the fluctuations
around a circular configuration, fulfilling eq. (2.8), in an arbitrary
spacetime of the form (2.3) (or (2.5)). As special cases we considered
Minkowski, de Sitter and arbitrary power law expanding
universes. In de Sitter space
we found that for {\it all} expanding strings the ratio of the mean fluctuation
amplitude to the radius of the unperturbed circular string is constant. In
power law inflationary universes, on the other hand, we found special
expanding solutions becoming more and more "circular".

In principle it is straightforward to consider any other FRW spacetime: first
one has to solve eq. (2.8), or at least find some special solutions, and then
the linear equations (2.22)-(2.23) have to be analyzed. Our formalism can
of course easily be generalized to describe fluctuations around
spherical membranes also. In fact, eq. (2.16) holds for an arbitrary defect
of arbitrary dimension embedded in an arbitrary curved spacetime of
higher dimension.
\newpage
\section{Appendix}
\setcounter{equation}{0}
In this appendix we give the explicit expressions for the non-vanishing
components of the Christoffel symbol and Riemann tensor, corresponding to the
line element (2.5).
\vskip 12pt
\hspace*{-6mm}{\bf The metric:}
\begin{equation}
g_{tt}=-(1-H^2f^2),\;\;\;g_{tf}=-Hf,\;\;\;g_{ff}=1,\;\;\;g_{\theta\theta}=
f^2,\;\;\;g_{\phi\phi}=f^2\sin^2\theta.
\end{equation}
\vskip 12pt
\hspace*{-6mm}{\bf The Christoffel symbol:}
\begin{eqnarray}
&\Gamma^t_{tt}=H^3f^2,\;\;\;\Gamma^t_{tf}=-H^2f,\;\;\;\Gamma^t_{ff}=H,\;\;\;
\Gamma^f_{ff}=H^2f,\;\;\;\Gamma^f_{tf}=-H^3f^2,&\nonumber\\
&\Gamma^f_{tt}=-f[H_{,t}+H^2(1-H^2f^2)],\;\;\;\Gamma^t_{\phi\phi}=
\sin^2\theta\Gamma^t_{\theta\theta}=Hf^2\sin^2\theta,&\nonumber\\
&\Gamma^f_{\phi\phi}=
\sin^2\theta\Gamma^f_{\theta\theta}=-f(1-H^2f^2)\sin^2
\theta,\;\;\;\Gamma^\theta_{f\theta}=
\Gamma^\phi_{f\phi}=\frac{1}{f},&\nonumber\\
&\Gamma^\theta_
{\phi\phi}=-\sin\theta\cos\theta,\;\;\;\Gamma^\phi_{\theta\phi}=\cot\theta.&
\end{eqnarray}
\vskip 12pt
\hspace*{-6mm}{\bf The Riemann tensor:}
\begin{eqnarray}
&R_{tftf}=-H_{,t}-H^2,\;\;\;R_
{t\phi t\phi}=\sin^2\theta R_{t\theta t\theta}=
f^2(H^4f^2-H^2-H_{,t})\sin^2\theta,&\nonumber\\
&R_{\theta\phi\theta\phi}=H^2f^4\sin^2
\theta,\;\;\;R_{t\phi f\phi}=\sin^2\theta R_{t\theta f\theta}=-H^3f^3
\sin^2\theta,&\nonumber\\
&R_{f\phi f\phi}=\sin^2\theta R_{f\theta f\theta}=H^2f^2\sin^2\theta.&
\end{eqnarray}
\vskip 12pt
\hspace*{-6mm}{\bf The Ricci tensor:}
\begin{eqnarray}
&R_{tt}=(H^2f^2-3)H_{,t}-3H^2(1-H^2f^2),\;\;\;R_{tf}=-Hf(H_{,t}+3H^2),&
\nonumber\\
&R_{ff}=H_{,t}+3H^2,\;\;\;R_{\phi\phi}=\sin^2\theta R_{\theta\theta}=
f^2(H_{,t}+3H^2)\sin^2\theta.&
\end{eqnarray}
\vskip 12pt
\hspace*{-6mm}{\bf The scalar curvature:}
\begin{equation}
R=6(H_{,t}+2H^2).
\end{equation}


\newpage
\begin{thebibliography}{11}
\bibitem{kib}E.W. Kolb and M.S. Turner, The early Universe (Addison-Wesley,
             Reading, MA, 1989).
\bibitem{bas}R. Basu, A.H. Guth and A. Vilenkin, Phys. Rev. {\bf D44}
             340 (1991).
\bibitem{san}C.O. Loust\'{o} and N. S\'{a}nchez, Phys. Rev. {\bf D47}
             4498 (1993).
\bibitem{all}A.L. Larsen, DEMIRM MEUDON 93/052, submitted
             to Phys. Rev. D.
\bibitem{gar1}J. Garriga and A. Vilenkin, Phys. Rev. {\bf D44} 1007 (1991).
\bibitem{gar2}J. Garriga and A. Vilenkin, Phys. Rev. {\bf D47} 3265 (1993).
\bibitem{gar3}J. Garriga and A. Vilenkin, Phys. Rev. {\bf D45} 3469 (1992).
\bibitem{mis}C.W. Misner, K.S. Thorne and J.A. Wheeler, Gravitation
             (Freeman, San Francisco, CA, 1973).
\bibitem{akd}K.G. Akdeniz, S. Erbil, H. Mutus and E. Rizaoglu, Phys. Lett.
             {\bf B237} 192 (1990).
\bibitem{li}X. Li and J. Zhang, Phys. Lett. {\bf B312} 62 (1993).
\bibitem{axe}A.L. Larsen and M. Axenides, Phys. Lett. {\bf B318} 47 (1993).
\bibitem{fro}A.L. Larsen and V.P. Frolov, Nucl. Phys. {\bf B414} 129 (1994).
\bibitem{car}B. Carter, Phys. Rev. {\bf D48} 4835 (1993).
\bibitem{guv}J. Guven, Phys. Rev. {\bf D48} 5562 (1993).
\bibitem{eis}L.P. Eisenhart, Riemannian Geometry (Princeton University Press,
             fifth printing, 1964).
\bibitem{vil}A. Vilenkin, Phys. Rev. {\bf D24} 2082 (1981).
\bibitem{san2}H.J. de Vega, A.L. Larsen and N. S\'{a}nchez, DEMIRM MEUDON
              93/055, LPTHE 93/56, submitted to Nucl. Phys.
\bibitem{mik}H.J. de Vega, A.V. Mikhailov and N. S\'{a}nchez, Teor. Mat. Fiz.
             {\bf 94} 232 (1993).
\end{thebibliography}
\end{document}